\documentstyle[multicol,prb,aps,epsfig]{revtex}

\renewcommand{\narrowtext}{\begin{multicols}{2}

\global\columnwidth20.5pc\noindent}
\renewcommand{\widetext}{\end{multicols}
\global\columnwidth42.5pc}
\begin{document}
\draft
\title{ Antiferromagnetic Zigzag Spin Chain in Magnetic Fields at Finite Temperatures}
\author{Nobuya Maeshima and Kouichi Okunishi${}^1$}
\address{Department of Physics, Graduate School of Science, Osaka University, Toyonaka, Osaka 560-0043, Japan.\\
${}^1$Department of Applied Physics, Graduate School of Engineering, Osaka University,Suita, Osaka 565-0871, Japan.\\
}
\date{\today}
\maketitle

\begin{abstract}
We study thermodynamic behaviors of the antiferromagnetic zigzag spin chain in magnetic fields, using the density-matrix renormalization group method for the quantum transfer matrix.
We focus on the thermodynamics of the system near the critical fields in the ground-state magnetization process($M$-$H$ curve): the saturation field, the lower critical field associated with excitation gap, and the field at the middle-field cusp singularity.
We calculate magnetization, susceptibility and specific heat of the zigzag chain in magnetic fields at finite temperatures, and then discuss how the calculated quantities reflect the low-lying excitations of the system related with the critical behaviors in the $M$-$H$ curve. 
\end{abstract}

\narrowtext

\section{introduction}

Magnetization processes ( $M$-$H$ curves, where $M$ is magnetization and $H$ is
magnetic field) of low-dimensional antiferromagnetic (AF) quantum spin systems have attracted much attention, in accordance with remarkable developments in material synthesis techniques and high-field experiments.
Recently, a lot of theoretical and experimental researches have clarified the mechanisms of various interesting behaviors of the $M$-$H$ curves at the zero temperature:  
e.g. the critical phenomena $\Delta M\sim \sqrt{H-H_{\rm c}}$ associated with the gapped  excitation  (excitation gap $\propto H_{\rm c}$)\cite{halgap,Aff1,tsv,SaTa1,OHA} or with the saturated magnetization (at the saturation field $H_{\rm s}$),\cite{SaTa1,OHA,HoP} magnetization plateau,\cite{s1alt1,platone,totsuka1,oshiyama} and, the middle-field cusp singularity(MFCS)\cite{Parkinson,SpinAlt,IntegLadder3,ohacusp}.
These are field-induced phase transitions of the ground states, reflecting the non-trivial energy level structures of the systems. 

Such non-trivial structures of the excitation often cause various characteristic behaviors on bulk quantities at finite temperatures as well. 
For instance, thermodynamic quantities in magnetic fields are calculated for ladder
systems,\cite{2leglad} mixed spin systems,\cite{ferri} etc, where various peak-structures of the specific heat are observed.

In this paper, we study the thermal behaviors of the antiferromagnetic zigzag spin chain in  magnetic fields, which is one of typical systems exhibiting the phase-transition behaviors in the $M$-$H$ curves and  actually realized as quasi-one-dimensional materials: SrCuO$_2$\cite{matsuda}, Cu(ampy)Br$_2$\cite{kikuchi} and the organic compound F$_2$PIMNH\cite{hosokoshi}.
However, the finite-temperature properties of the zigzag chain in the thermodynamic limit have not been studied quantitatively so much,\cite{harada,maisinger,klumper,tota} since the frustrated interaction makes the reliable Quantum Monte Carlo(QMC) simulation difficult\cite{nsp}. 
Thus it is a fairly interesting problem to investigate the finite-temperature behavior of the zigzag chain both from theoretical and experimental view points.  

The Hamiltonian of the zigzag chain is given by
\begin{eqnarray}
{\cal H}&=& J \sum_{i} [\vec{S}_{i}\cdot\vec{S}_{i+1}
+j  \vec{S}_{i}\cdot\vec{S}_{i+2}] 
- g \mu_B H \sum_i S_i^z, \label{ladder}
\end{eqnarray}
where $\vec{S}_i$ is the $S=1/2$ spin operator at $i$-th site, $g$ is the
$g$-factor, and $\mu_B$ is the Bohr magneton.
We have denoted the nearest neighbor coupling as $J$,  the ratio of the next-nearest coupling as $j$($>0$), and  the applied field as  $H$.
In ref. 14, 
it has been shown that the zigzag chain has fascinating $M$-$H$ curves as varying $j$. 
Near the saturation field, there exists the MFCS for $j>1/4$ , and the associated two component Tomonaga-Luttinger liquid behavior is observed, where the dispersion curve of the elementary excitation is the double-well curve. 
At just $j=1/4$, the $M$-$H$ curve  behaves  as $\Delta M \sim (H-H_{\rm s})^{1/4}$, unlike the usual square-root behavior in $j<1/4$.\cite{unusual} 
Near the zero field, the  $M$-$H$ curve in the spin-fluid phase $j<j_{\rm fd}(\equiv 0.2411)$\cite{nomura} is similar to that of the $S=1/2$ Heisenberg spin chain.
For $j>j_{\rm fd}$, the $M$-$H$ curve exhibits the square-root behavior, since the system is in the dimerized phase and becomes gapful. 
Further, for $j>0.5$, the incommensurate behavior of the ground state  correlation function\cite{haratone,whiteaffleck,watanabe} suggests that another MFCS appears in the $M$-$H$ curve near the lower critical field.  
We focus on how the above characteristic behaviors of the $M$-$H$ curves influence the thermodynamic properties of the zigzag chain.  

In order to calculate the finite-temperature quantities of the zigzag chain, we employ the density matrix renormalization group (DMRG) method\cite{DMRG} for the quantum transfer matrix(QTM).\cite{wang,shibata}
The remarkable point is that the DMRG is free from the negative sign problem and thus it is successfully applied to some frustrated spin ladder systems.\cite{maisinger,klumper}
We calculate the magnetization $M$, susceptibility $\chi$, and specific heat $C$  at various temperatures down to $T/J\sim 0.05$ in the thermodynamic limit. 
Then we discuss the effects of the non-trivial energy spectrum on the obtained quantities.

This paper is organized as follows.
In section 2, we describe the DMRG for the QTM in brief.
In section 3, we show the calculated results for the $M$-$H$ curves of the zigzag chain.
In section 4, the thermodynamic properties of the zigzag chain near the saturation field are discussed in detail.
In particular, we consider the effect of the MFCS on the specific heat.
In section 5, we calculate the quantities $\chi$, $C$, and $C/T$ near the zero field, and discuss the relation to the low-energy excitation of the system. 
Conclusions are summarized in section 6.


\section{numerical methods}

The DMRG method for the QTM is widely used  to study one-dimensional(1D) quantum spin systems at finite temperatures numerically,\cite{wang,shibata} 
since the physical quantities can be obtained successfully down to low temperatures, where the QTM with a sufficient large Trotter number should be treated accurately.
Further, it should be mentioned that the DMRG for the QTM is free from the negative sign problem the QMC simulation is suffering from.
Thus we can say that the DMRG for the QTM is one of the most suitable numerical tools to study the zigzag chain.

To see our strategy  concretely,  we rewrite the Hamiltonian of the zigzag chain with the lattice length $2L$ into a ladder form shown in Fig. 1:
\begin{eqnarray}
{\cal H}=J\sum_{i=1}^{L} \hat{h}_{i,i+1} ,
\end{eqnarray}
with
\begin{eqnarray}
\hat{h}_{i,i+1}&=& \frac{1}{2}[\vec{S}^{A}_{i}\cdot\vec{S}^{B}_{i}+ \vec{S}^{A}_{i+1}\cdot\vec{S}^{B}_{i+1}] 
+ \vec{S}^{B}_{i}\cdot\vec{S}^{A}_{i+1} \nonumber  \\
&+& j[\vec{S}^{A}_{i}\cdot\vec{S}^{A}_{i+1}+\vec{S}^{B}_{i}\cdot\vec{S}^{B}_{i+1}] + \; {\rm Zeeman \; terms},
\end{eqnarray}
where $A$ and $B$ are the labels of the lower and upper legs respectively.

\begin{figure}
\epsfxsize=2.6 in\centerline{\epsffile{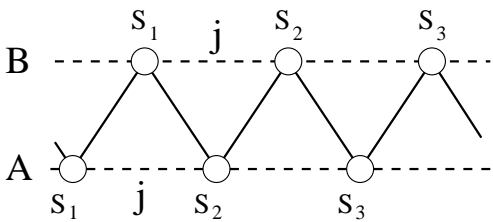}}
\caption{The ladder representation of the zigzag spin chain}
\label{zigham}
\end{figure}

By using the Suzuki-Trotter decomposition,  we map the zigzag chain at finite temperature into a 2D classical system on a checker board lattice.\cite{trotterdcmp} 
Then the partition function of the system can be represented by the QTM $T^{(N)}_{\rm e}T^{(N)}_{\rm o}$:
\begin{equation} 
 Z= \lim_{N\to\infty}{\rm Tr} [(T^{(N)}_{\rm e}T^{(N)}_{\rm o})^{L/2} ],
\end{equation}
where $N$ is the Trotter number. 
The QTM is defined by the product of the local operator $W_{i,k}$ originating from the local Boltzmann weight in the mapped system: 
\begin{eqnarray}
T_{\rm e}^{(N)}&=&\prod_{2 \le k \le 2N;\; k\in {\rm even}} W_{i,k} , \quad {\rm for} \quad i={\rm even} \\
T_{\rm o}^{(N)}&=&\prod_{1 \le k \le 2N-1;\; k\in {\rm odd}\;} W_{i,k} , \quad {\rm for}\quad  i={\rm odd}
\end{eqnarray}
where $i$ is the index of the spatial direction and $k$ is that of the Trotter one.
By using the $S^z$-diagonal representation $|s\rangle$ of the spin operator $\vec{S}$, we have the explicit element of the weight:
\begin{eqnarray}
 &{}&\langle s^{A,B}_{i+1,k} s^{A,B}_{i+1,k+1}|W_{i,k}|s^{A,B}_{i,k}s^{A,B}_{i,k+1}\rangle  = \nonumber \\
 &{}& \quad \langle s^{A,B}_{i,k+1}  s^{A,B}_{i+1,k+1}  |\exp(-\epsilon\hat{h}_{i,i+1})|s^{A,B}_{i,k} s^{A,B}_{i+1,k} \rangle,
\end{eqnarray}
where $\epsilon=J/(k_BT N)$ and the notation $| s^{A,B}_{i,k}\rangle \equiv |s^A_{i,k}, s^B_{i,k}\rangle$ is introduced for simplicity.

In the thermodynamic limit $L\to \infty$, the thermal property of the system is
extracted from the largest eigenvalue and the corresponding eigenvector of the QTM with the sufficiently large Trotter number $N$. 
Then we can employ the DMRG for the transfer matrix developed by T.Nishino.\cite{nishino}
Practically, we regard, for example,  $| s^{A,B}_{i,k}\rangle$ as a four-state single spin, and  perform the DMRG calculation with the periodic boundary condition. 
The detail of the algorithm follows  Ref.35, where $T^{(N)}_{\rm e}$ and $T^{(N)}_{\rm o}$ are renormalized separately.
In addition we make the density matrix block-diagonal with help of the conservation law for the QTM.\cite{wang} 
 
In the DMRG calculation, we often meet undesirable complex eigenvalues of the density matrix, because the numerical diagonalization yields inaccurate results for an asymmetric matrix having degenerate eigenvalues.
Thus, if the complex number appears in the nearly-degenerate eigenvectors of the density matrix, we reorthogonalize the corresponding eigenvectors to be represented by real numbers.\cite{eggrom}

We calculate the magnetization $M$ and the internal energy $E$  from the obtained eigenvector directly. 
We further calculate the susceptibility $\chi$ and specific heat $C$ by numerical differentiation for $M$ and $E$ respectively. 
In the following,  we set $k_B=1$,  $g\mu_B=1$ and $J =1$ for simplicity.
The calculations were done with $\epsilon=0.1$ and the maximum number of the retained bases $m=88$. We have confirmed that the computed data converged with respect to $m$ and $\epsilon$.


\section{numerical results for the $M$-$H$ curves}

\begin{figure}
\epsfxsize=2.8 in\centerline{\epsffile{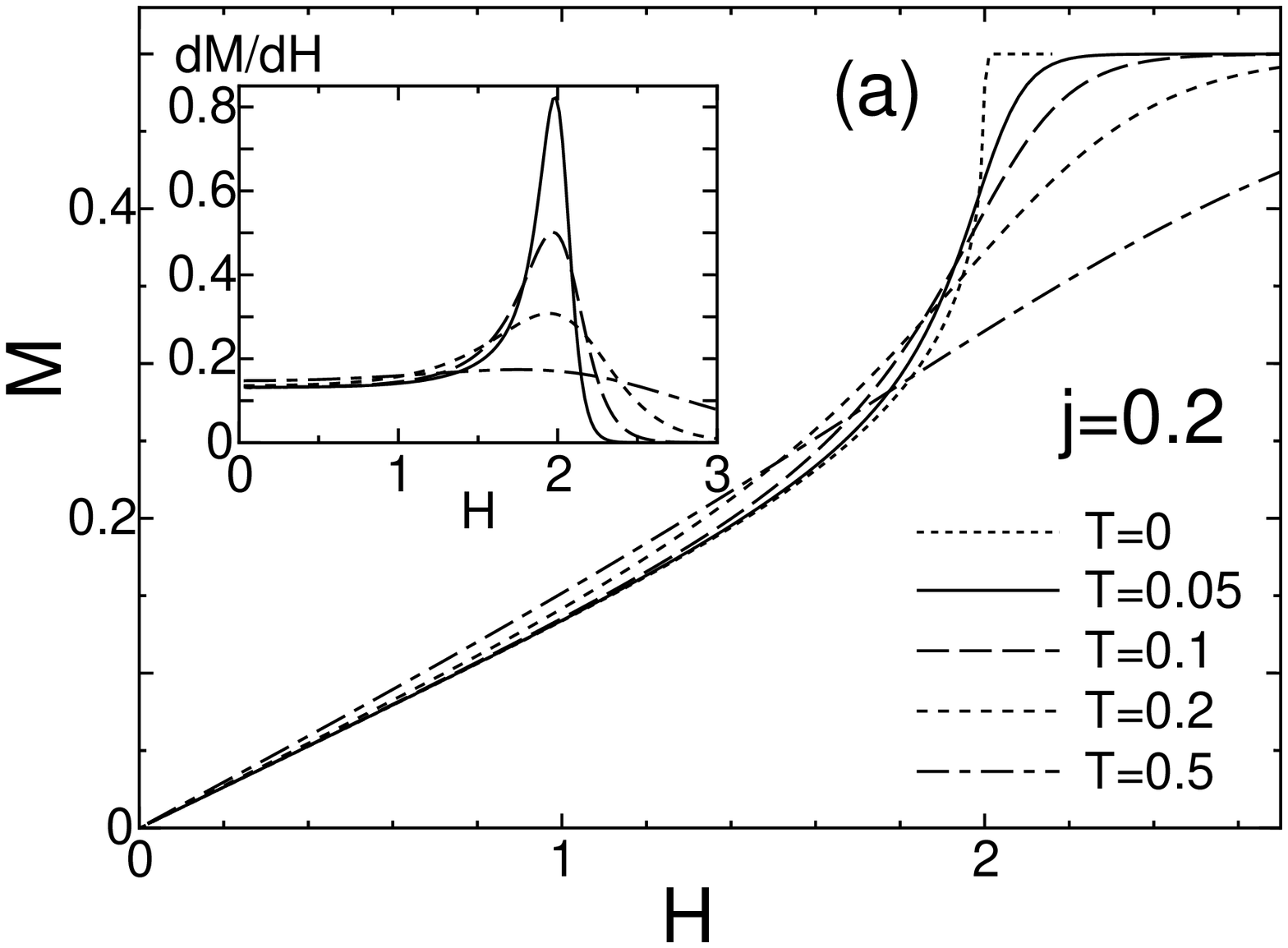}}
\epsfxsize=2.8 in\centerline{\epsffile{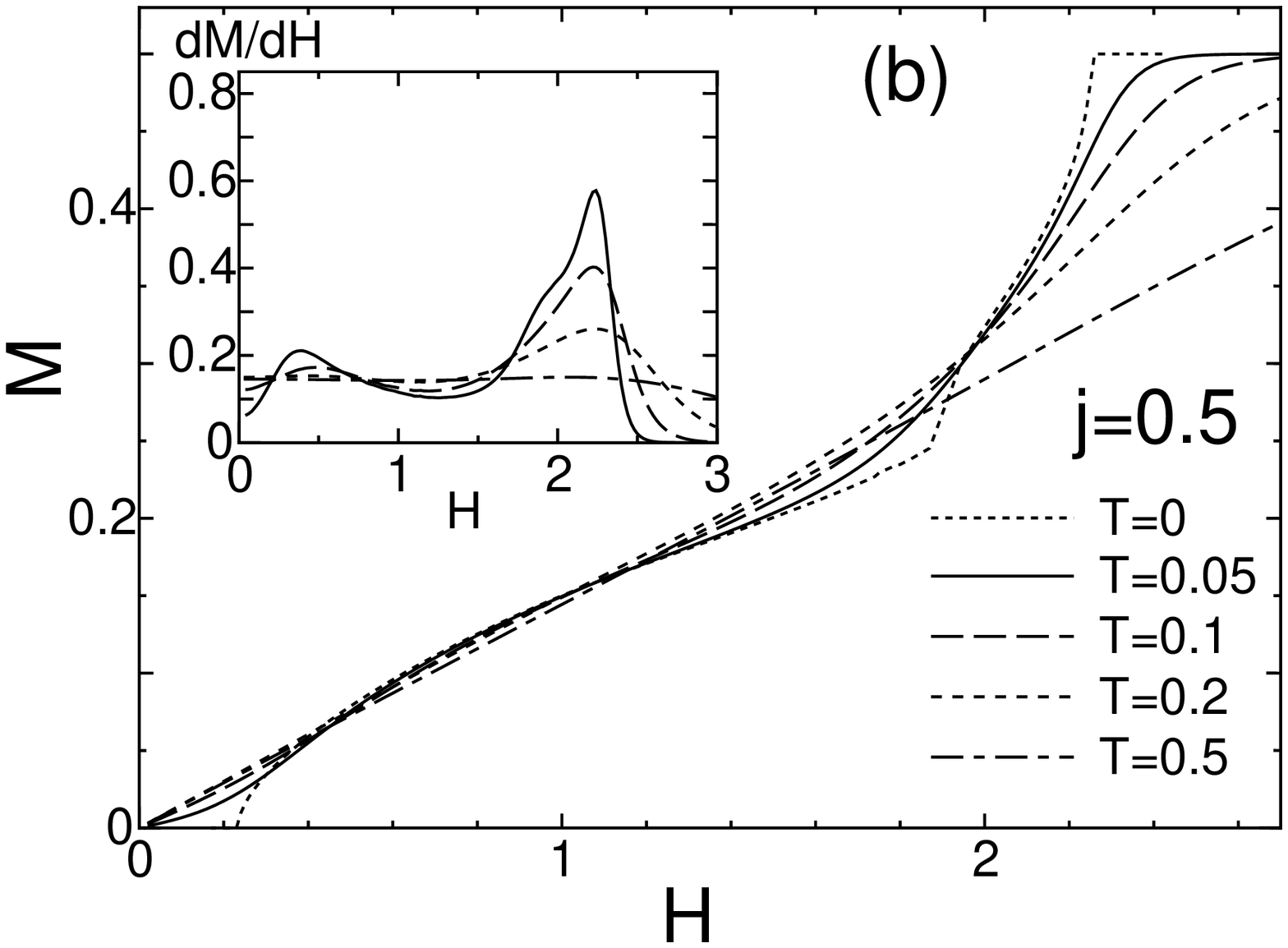}}
\epsfxsize=2.8 in\centerline{\epsffile{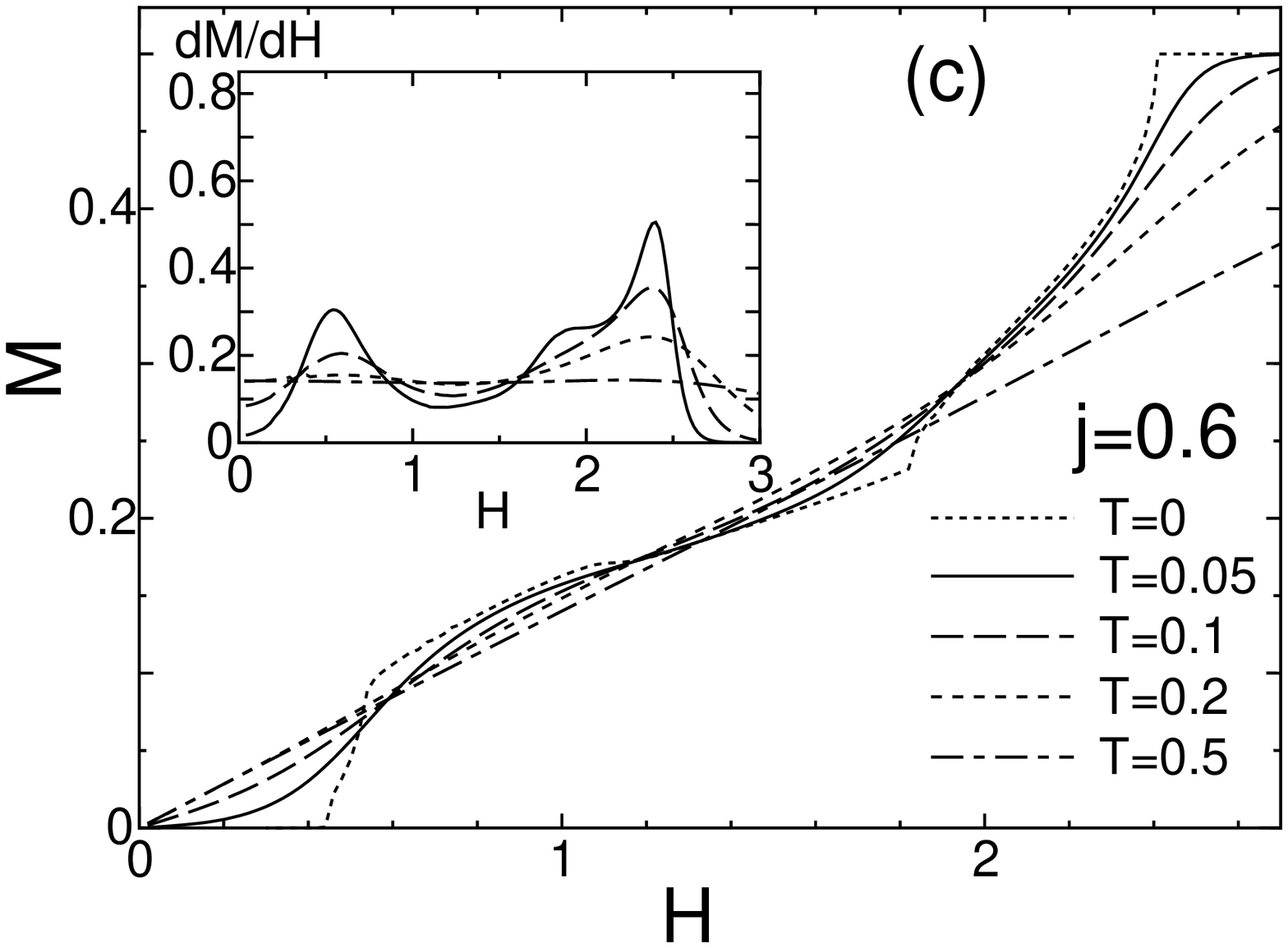}}
\caption{ the $M$-$H$ curves of the zigzag chains at finite temperatures.
(a) $j=0.2$, (b) $j=0.5$, and (c) $j=0.6$.
Inset: the differentiation of the $M$-$H$ curve ($dM/dH$ curve). }
\label{zigzagmh}
\end{figure}

In Fig. \ref{zigzagmh}, we show the calculated  $M$-$H$ curves of the zigzag chain at finite temperatures for $j=0.2$, $0.5$, and $0.6$. 
We also show the zero-temperature $M$-$H$ curves calculated by the product-wavefunction renormalization group method (PWFRG)\cite{PWFRG} for comparison.
At zero temperature, the $M$-$H$ curve for $j=0.2$ has no anomaly in the middle field region. 
On the other hand, the $M$-$H$ curve for $j=0.5$ has one MFCS at $H_{\rm cusp} \simeq 1.9$. 
We  further find that  the $M$-$H$ curve for $j=0.6$ has two MFCS at $H_{\rm cusp}\simeq 0.6$ and $1.8$.

At finite temperature, the singularities of the ground state $M$-$H$ curve are
generally rounded by the thermal excitation.
Indeed  the $M$-$H$ curves show no singularity in the high temperature region.
As temperature decreases, however, the quantum effects appear.
We can see that the temperature dependence  of the $M$-$H$ curve for $j=0.5$ is enhanced near the cusp field $H_{\rm cusp} \simeq 1.9$, comparing with that for $j=0.2$.  
The differentiation curve($dM/dH$ curve) of the $M$-$H$ curve shows the MFCS effect more clearly(insets of Fig. \ref{zigzagmh}). 
In the $dM/dH$ curve for $j=0.5$ at $T=0.05$, a shoulder can be observed at $H=2.0$.
For $j=0.6$ we can see the winding structure in the $M$-$H$ curve at $T=0.05$,
corresponding to the higher MFCS and the lower MFCS. 
The $dM/dH$ curve also displays the double cusp structure of $M$-$H$ curve at low temperature;  the shoulder at the higher MFCS field becomes more clearly and the peak near the lower MFCS field does sharp.

\section{near the saturation field}

In this section, we consider the thermodynamic behaviors of the zigzag chain near the saturation field, where we can take a down-spin-particle picture of the elementary excitation.\cite{ohacusp} 
The  saturation field  $H_s$ is given by  $2$ for $j \le 1/4$ and $1+2 j +1/(8 j)$ for $j>1/4$. 
Near $H_s$, a down spin in the saturated (all up) state can be regarded as a spinless fermion, which is the effective low-energy limit of the $\delta$-function Bose gas model\cite{SaTa1,OHA}, and then the energy dispersion of the particle is calculated to be  
\begin{equation}
\omega(k)= \cos k-1 +j(\cos 2k-1).
\label{zzdisp}
\end{equation}
This one-particle dispersion curve fully  characterizes the qualitative property of the $M$-$H$ curve near $H_s$. 
For $j\le 1/4$, $\omega(k)$ has a single minimum at $k=\pi$, while, for $j>1/4$, $\omega(k)$ has a local maximum at $k=\pi$ and two minima at $k=\pi\pm\cos^{-1}(1/(4j))$.\cite{saturation,ohacusp} 
Thus, for $0\le j \le 1/4$, the $M$-$H$ curve is smooth in the whole field range $0 \le H \le H_s$.
While, for $j>1/4$, `` van Hove singularity'' corresponding to the local maximum of $\omega(k)$ gives the MFCS in the higher-field region of the $M$-$H$ curve.\cite{ohacusp}

\begin{figure}
\epsfxsize=2.8 in\centerline{\epsffile{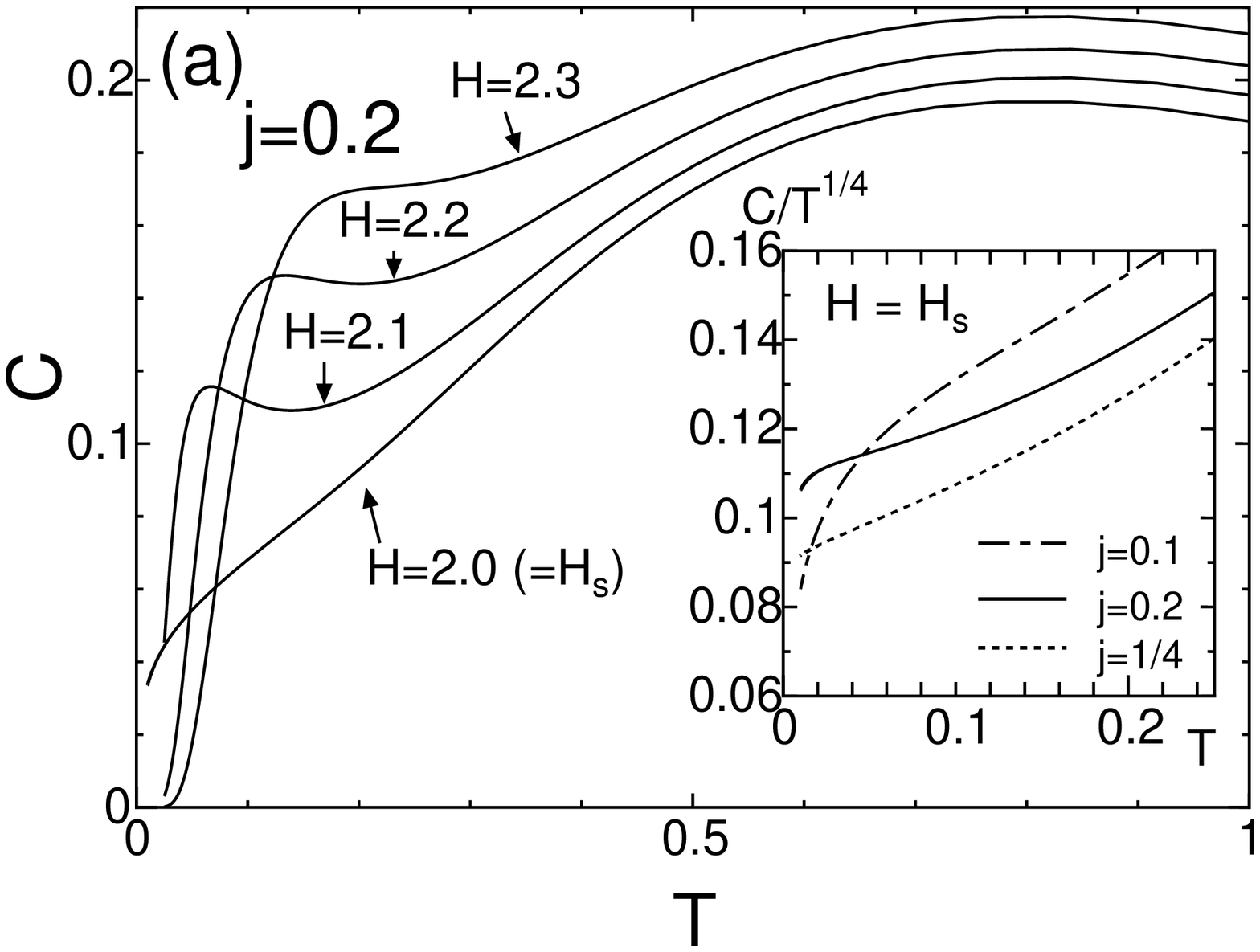}}
\epsfxsize=2.8 in\centerline{\epsffile{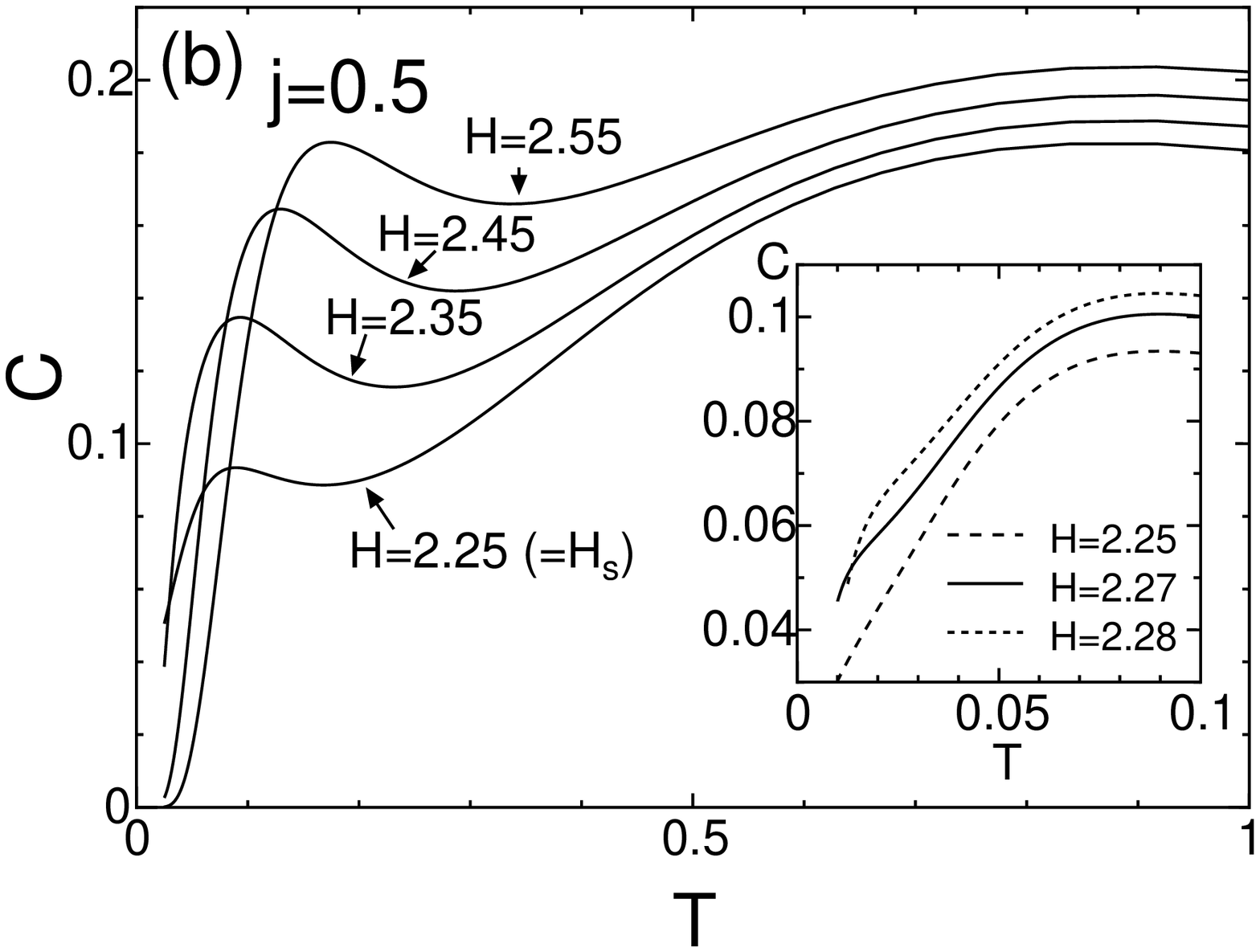}}
\caption{ The specific heat $C$ near $H=H_s$.  (a) $j=0.2$ and (b) $j=0.5$. 
Inset of figure (a): $C/T^{1/4}$ plot for $j=0.1$, $0.2$ and $j=1/4$ at $H=H_s$.
Inset of figure (b): magnification in low-temperature region at the slightly upper
fields than $H_s$. }
\label{sp1}
\end{figure}

In Fig. \ref{sp1}, we show the specific heat $C$ at various applied fields $H$ above $H_s$. 
In the case of $j=0.2$,  clear peaks of $C$ can be seen in the low-temperature region($T<0.3$), except for $H=H_s$.
Then we can find that the peak position $T_{\rm peak}$, which is defined by the temperature at the top of the low-temperature peak, is approximately proportional to the energy gap $|H - H_s|$.  
Hence, the low-temperature peaks represent the divergence of the density of state at the bottom of the dispersion curve (\ref{zzdisp}).

At just $H=H_s$, the peak disappears and, then, the temperature dependence of the specific heat $C(H=H_s)$ becomes the power-law behavior: $T^{1/\alpha}$ in $T\ll 1$(see the inset of Fig. \ref{sp1}(a) ).
This power $\alpha$ is determined by the shape of the dispersion curve in $k\to \pi$
limit, where we have $\omega(k)\sim (k-\pi)^\alpha$.
For $j<1/4$, in strict sense, $\alpha=2$ should be obtained in $T\to 0$ limit, since Eq. (\ref{zzdisp}) has a quadratic term. 
However, the coefficient of the quadratic term decreases and the quartic term becomes dominant, as $j$ increases to $j=1/4$.
Thus we see the cross-over behavior from $\alpha=4$ to $2$ as shown in the inset of
Fig. 3(a). On the other hand, $C/T^{1/4}$ plot for $j=1/4$ is consistent with
$\alpha=4$, where the quadratic  term vanishes completely and Eq. (\ref{zzdisp}) becomes $\omega (k) \sim (k-\pi)^4$.
The above power-law behavior of $C$ is essentially the same as the Fermi liquid one, which is verified for the Heisenberg chain($j=0$) with $\alpha=2$ \cite{xiang}.

For $j=0.5$,  the double-well structure of Eq. (\ref{zzdisp}) induces interesting properties on the specific heat $C$ near $H_s$. 
Indeed, the low-temperature peak exists  even at $H=H_s (=2.25)$, and a weak shoulder structure can be observed at the slightly upper fields than $H_s$ (see the Inset of  Fig. \ref{sp1}(b) ). 
In addition, we can see that the peak-temperature $T_{\rm peak}$ is approximately proportional to $H - H_{\rm cusp}$ when $|H_s-H|\ll 1$.  
Therefore the strong peak represents the ``van Hove singularity'' corresponding to the MFCS, and the weak shoulder structure comes from the bottom of the dispersion curve.

\begin{figure}
\epsfxsize=2.7 in\centerline{\epsffile{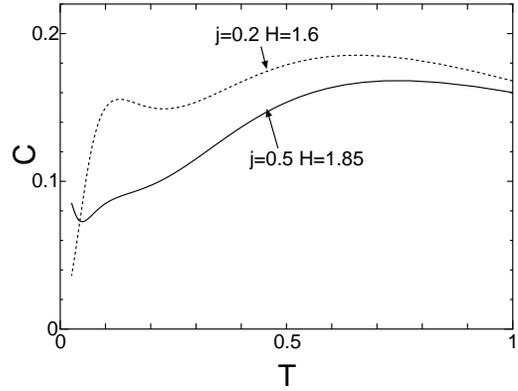}}
\caption{ The specific heat $C$ near $H=H_{\rm cusp}$. For comparison, we show that for $j=0.2$ at $H=1.6$.
 }
\label{sp2}
\end{figure}

In order to illustrate the effects of the MFCS more clearly, we have calculated the specific heat $C$ near the cusp field $H_{\rm cusp}\simeq 1.9$. 
In Fig. \ref{sp2}, we show the calculated result for $j=0.5$ at $H=1.85 (=H_s-0.4)$, which is slightly below $H_{\rm cusp}$. 
We also show that for $j=0.2$ at $H=1.6 (=H_s-0.4)$ for comparison.
For $j=0.5 $, we can find that the specific heat increases sharply at very low
temperature, and has a shoulder at $T\sim 0.1$. 
On the other hand, for $j=0.2$, we find a peak near $T\sim 0.1$ only.
Thus, the sharp increase of the specific heat at very low temperature for $j=0.5$ comes from the ``van Hove singularity'' of $\omega(k)$, and the shoulder is attributed to the bottom of $\omega(k)$.  
In addition, we note that the bottom of $\omega(k)$ gives weak contribution to the specific heat. 

From the above results,  we can conclude that the band edge singularities of $\omega(k )$ at $k=\pi$ or  $k=\pi\pm\cos^{-1}(1/4j)$ explain correctly the specific heat behaviors both near $H_s$ and $H_{\rm cusp}$.

\section{Near the zero applied  field}

At the zero magnetic field,  various experiments have been done for the actual materials\cite{matsuda,kikuchi,hosokoshi}, where  the observed susceptibility is shifted from that of the pure $S=1/2$ Heisenberg model.
This shift of the susceptibility is attributed to the deformation of the low-lying excitation, which is induced by the frustration effect due to the next-nearest coupling.
In this section, we calculate the thermodynamic quantities near the zero field, which play a crucial role to determine the exchange coupling constant of the materials, and discuss the relation between the obtained quantities and the low-lying excitation structures.

\begin{figure}
\epsfxsize=2.8 in\centerline{\epsffile{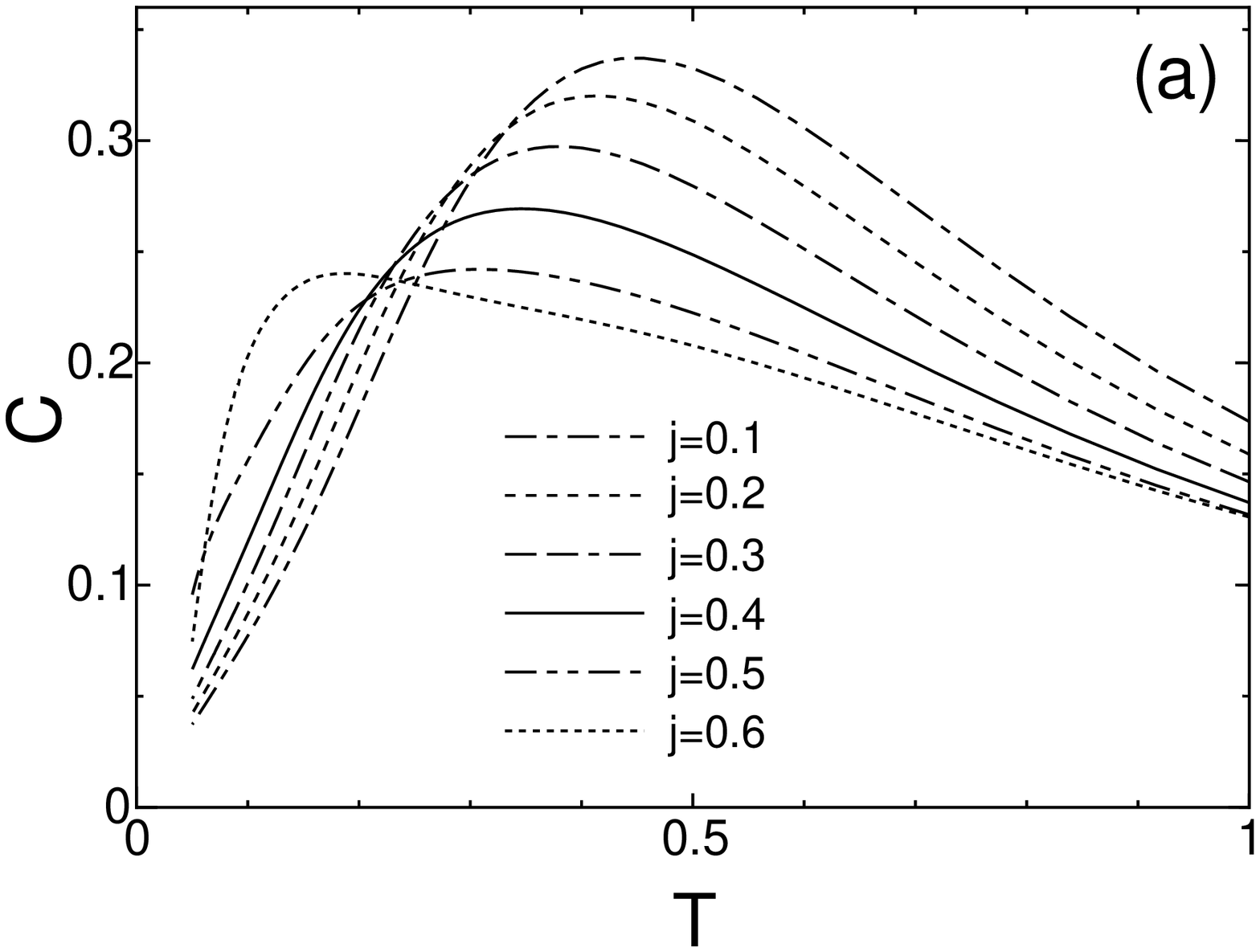}}
\epsfxsize=2.8 in\centerline{\epsffile{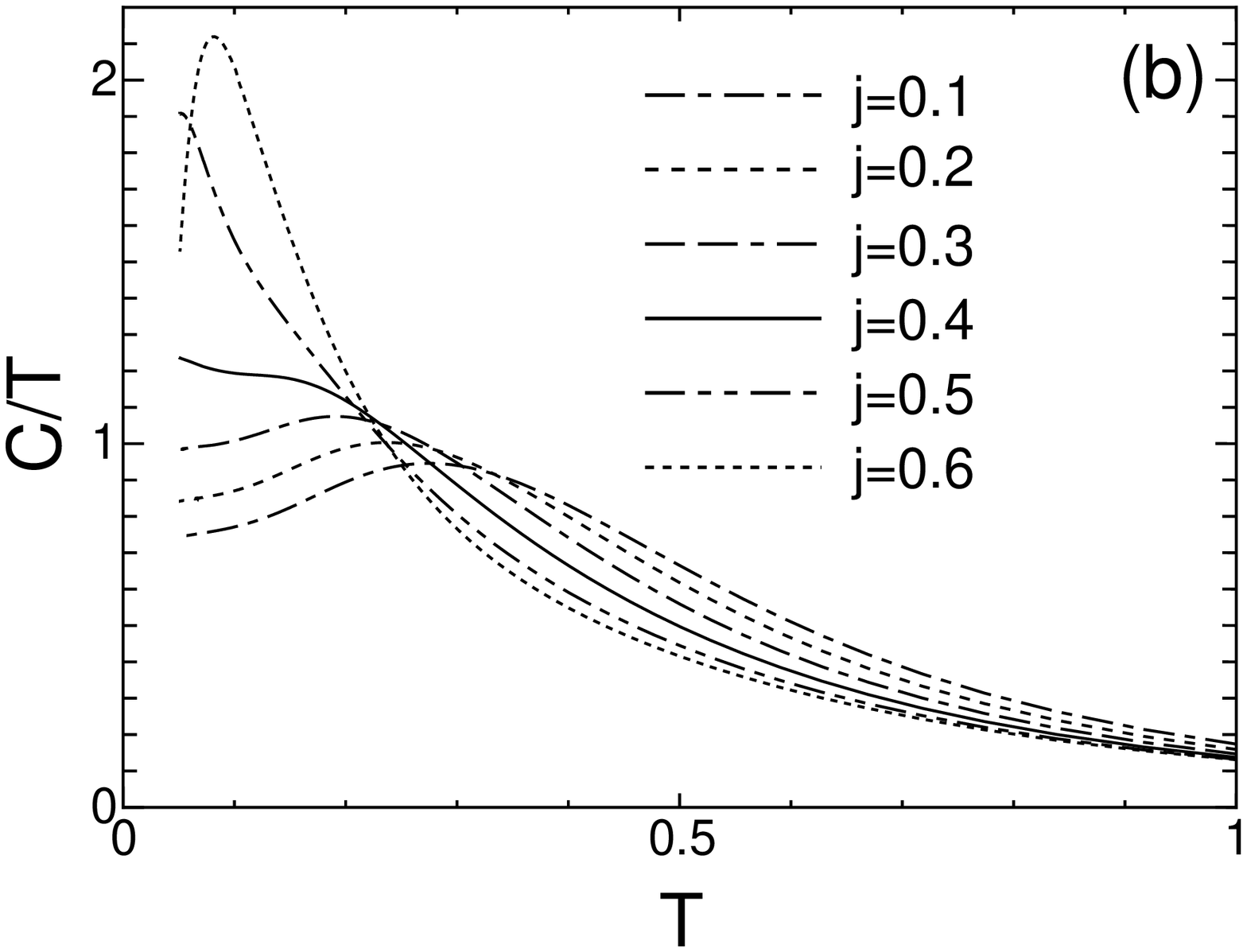}}
\epsfxsize=2.8 in\centerline{\epsffile{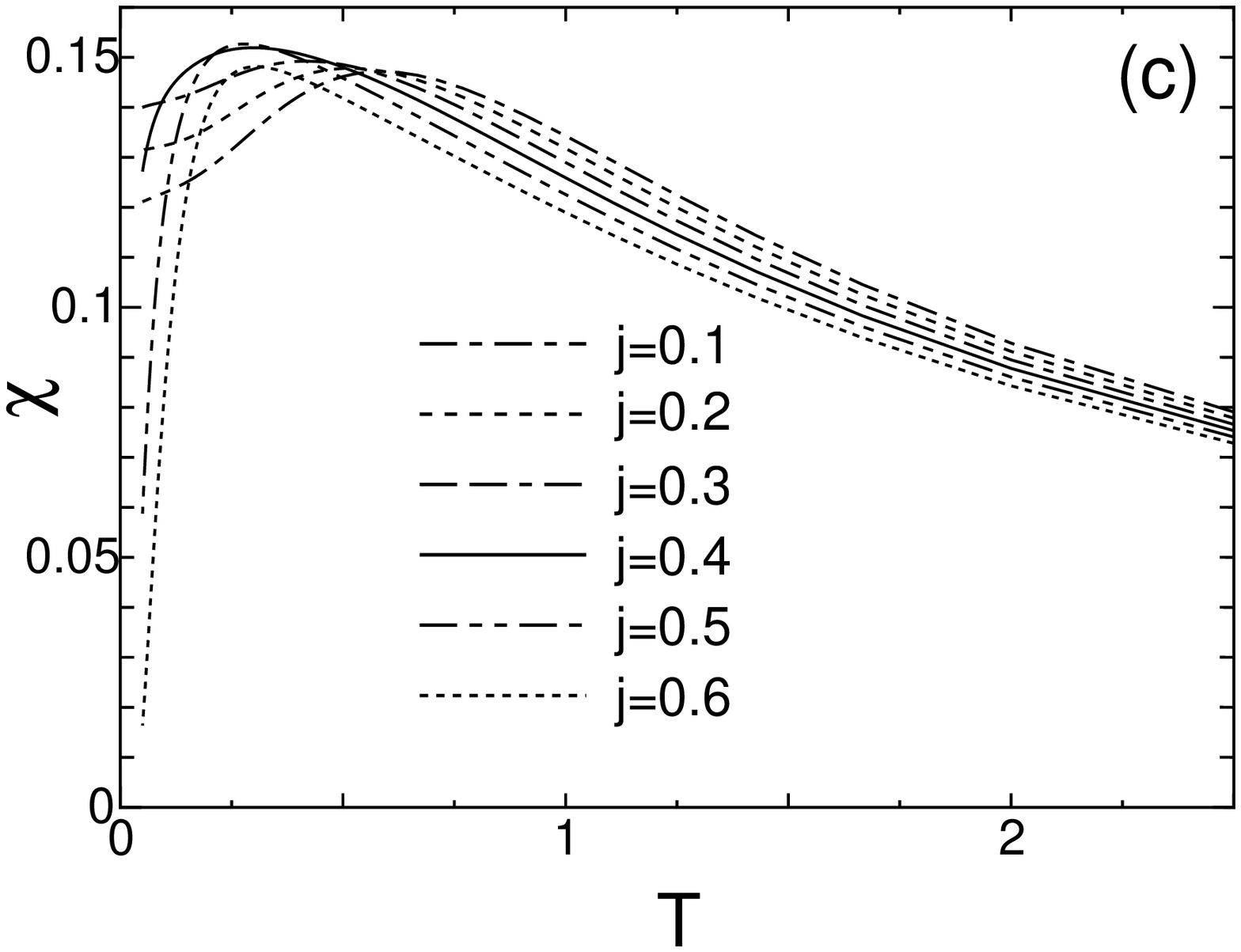}}
\caption{Thermodynamic quantities at zero magnetic field
 (a) The specific heat $C$, (b) $C/T$, and  (c) The susceptibility $\chi$. }
\label{sus}
\end{figure}

In Fig. \ref{sus}, we show the specific heat $C$, the quantity $C/T$, and the
susceptibility $\chi$  at the zero field for $0.1\le j \le 0.6$. 

In $j<j_{\rm fd}(=0.2411)$, the zigzag chain is in the gapless spin-fluid phase at
the zero field and the $M$-$H$ curves do not show any singularity as in
Fig. \ref{zigzagmh}(a).
Thus the calculated quantities for $j=0.1$ and $j=0.2$ are qualitatively the same as that of  the $S=1/2$ Heisenberg chain($j=0$), though the broad peaks of $\chi$ and  $C$ shift to the low-temperature side.
As shown in Fig. \ref{sus}(a) and (b),  $\chi$ and $C/T$ approaches to the constant value in the zero-temperature limit.
We extrapolate the coefficient of the linear specific heat $\gamma\equiv \lim_{T \to 0} C/T$ by a polynomial fit of $C/T = \gamma + a_1 T + a_2 T^2$ in the range $0.06<T < 0.12$, and obtain $\gamma=0.75$  for $j=0.1$ and $\gamma=0.85$ for $j=0.2$. 
We also compute the zero-temperature susceptibility $\chi_0$ by numerical differentiation of the PWFRG results with $\Delta H=0.005$: we obtain $\chi_0=0.118$ for $j=0.1$ and $\chi_0=0.131$ for $j=0.2$.
Here we note that the same manner of the extrapolation at the Heisenberg point ($j=0$) yields  $\gamma=0.66$ and $\chi_0 = 0.108$, which  agree with the Bethe ansatz values.
According to the conformal field theory(CFT),  the coefficient of the linear specific heat and the ground state susceptibility $\chi_0$ are given by $\gamma=\pi c /(3v)$ and $\chi_0 =1/(2 \pi v)$ respectively, where $c$ is the central charge and $v$ is the spin wave velocity. 
The calculated  values of $\gamma$ and $\chi_0$ are almost consistent with the CFT prediction with $c=1$, but the strong log-correction for $\chi_0$\cite{nomura,griffis,eggart} prevents us from verifying the CFT relation precisely.

For $j>j_{\rm fd}$ the zigzag spin chain is gapful.
However, the magnitude of the gap is quite small in $\displaystyle j_{\rm fd}<j \mathop{<}_{\sim}0.4$.\cite{haratone,whiteaffleck}
Thus the specific heat $C$ and the susceptibility $\chi$ for $j=0.3$ are similar to those in $j<j_{\rm fd}$.
For $j=0.4$ and $ 0.5$, $\chi$ and $C$ catch the effect of the gap:
the  exponential decay of the susceptibility can be seen in the low temperature. 
For $j=0.6$, the energy gap becomes large, so that $\chi$ and $C$ show clear exponential decay in low temperature region.
In the specific heat for $j=0.6$, the peak at $T\simeq 0.15$ seems to be
distinguishable from  the broad peak around $T\simeq 0.5$(see  Fig. \ref{sus}(a)).
As shown in Fig. \ref{zigzagmh}(c), the $M$-$H$ curve of $j=0.6$ has the lower MFCS, and  the dispersion curve of the excitation is expected to be the double-well structure, which is supported by the incommensurability  of the ground-state correlation function.\cite{haratone,whiteaffleck,watanabe} 
The double-well structure of the dispersion curve possibly explains the remarkable shape change of the specific heat for $j=0.6$.

\begin{figure}
\epsfxsize=2.8 in\centerline{\epsffile{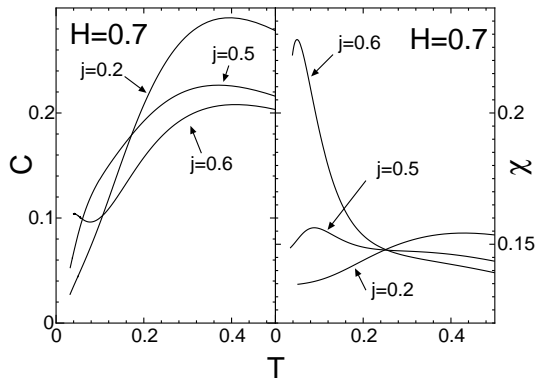}}
\caption{The specific heat $C$ and the susceptibility $\chi$ in the magnetic field $H=0.7$.} 
\label{mf07sp}
\end{figure}

Further, we consider the zigzag chain in a small magnetic field, where the  excitation related with the lower MFCS can be observed  more directly. 
In Fig. \ref{mf07sp}, we show the specific heat $C$ and the susceptibility $\chi$ for $j=0.2$, $0.5$,and  $0.6$ at the fixed field $H=0.7$.  
For $j=0.2$,  the properties of $C$ and $\chi$ is essentially the same as those at the zero field. 
For $j=0.5$, on the other hand,  $C$ and $\chi$ are enhanced in the low temperature region, because of the excitation gap.
For $j=0.6$,  where the lower MFCS appearing in the $M$-$H$ curve,  we can see the susceptibility has very sharp peak at very low temperature.
The specific heat also grows sharply at very low temperature as well, which is quite similar to that of the higher field MFCS.
These outstanding peaks in $\chi$ and $C$ support that the dispersion curve of the low-lying excitation has such a structure as the double-well one.
However, the evidence for the double-well structure can not be detected directly within the present calculation, since the field at the lower MFCS and the lower critical field of the excitation gap are very close as shown in Fig. \ref{zigzagmh}(c).

Here, we make a comment on the relevance of present calculations with the experimental results.
For example, the susceptibility of the compound Cu(ampy)Br$_2$ was measured by Kikuchi, and then the ratio of the exchange coupling is estimated as $j=0.2$. \cite{kikuchi}
The $M$-$H$ curve of this compound is also observed to have no anomaly in the middle field region, which is consistent with the present calculation.
Moreover, Hosokoshi, {\it et al} have found that the susceptibility of the organic compound F$_2$PIMNH shows the clear exponential decay in low temperature region\cite{hosokoshi}, which imply $j>0.4$ compared with the calculated susceptibility in  Fig. \ref{sus}-(c). Then it is  suggested that the $M$-$H$ curve of F$_2$PIMNH has the MFCS. However the high field experiment has not yet been done for this compound.

\section{summary}

In this paper we have quantitatively studied the thermodynamic behaviors of the antiferromagnetic zigzag spin chain in magnetic fields.
We have calculated the magnetization processes ($M$-$H$ curves) at finite temperatures by using the density-matrix renormalization group (DMRG) method for the quantum transfer matrix. 
The zero-temperature  $M$-$H$ curves were also calculated by the product wave-function renormalization group method.
We have shown that the  zero-temperature $M$-$H$ curve has  zero, one, and two
middle field cusp singularity(MFCS) for $j=0.2$, $0.5$ and $0.6$, respectively. The thermal effect on the $M$-$H$ curve was discussed as well.

We have further investigated the bulk physical quantities at finite temperatures in terms of the non-trivial energy-level structures responsible for the singularities in the zero-temperature $M$-$H$ curve.
Near the saturation field, we have considered the band edge singularities in $j\le1/4$, where the Fermi liquid behavior is observed.
For $j>1/4$, we have shown that the double-minimum shape of the dispersion curve  correctly explains the peak structures of the specific heat.

Near the zero field, we have considered the susceptibility $\chi$ and specific heat $C$.
In the gapless region, i.e. $j<0.2411$, we have shown  $\chi$ and $C$ exhibiting typical gapless behaviors  and checked the CFT relation between the zero-temperature susceptibility and the coefficient of linear specific heat.
For $j>0.2411$, the exponential decay of $\chi$ and $C$ has been observed in low temperatures.
Moreover, we have seen that another peak of $C$ is induced for $j=0.6$,  where the double-well structure of the dispersion curve is expected to accompany with the lower MFCS.
We have also calculated the susceptibility $\chi$ and specific heat $C$ in the small field $H=0.7$. 
The highly enhanced peaks for $\chi$ and $C$ of $j=0.6$ support the double-well
structure of the dispersion curve.
To understand the microscopic view about such low-lying excitation connected to the conformation mechanism of the lower MFCS is a remaining problem.

In the connection to the experiments,  the $M$-$H$ curve of F$_2$PIMNH is
interesting, since the appearance of the MFCS can be predicted by the present calculation.
In addition, the crystal structure of another zigzag materials\cite{hagiwara} suggests that they may have large next-nearest coupling $j$, by taking into account the orbital symmetry of the atoms concerned with the super-exchange interaction.
In analyzing experimental data for the above zigzag materials, we believe that the present results is of use.

\acknowledgments

We would like to thank H. Kikuchi, Y. Narumi, and M. Hagiwara for stimulating
discussions about experiments. 
We also thank to Y. Akutsu, Y. Hieida for fruitful discussions.
One of the authors (K. O.) is supported by the Japan Society for the Promotion of Science.

\widetext

\begin{references}
\bibitem{nsp} The elimination of the negative sign has been achieved for the limited cases.  See, for example, ref.\cite{tota}.
\end{references}
\end{document}